\documentclass[12pt, draftclsnofoot, onecolumn]{IEEEtran}
\usepackage{graphicx}
\usepackage{amsmath,bbm,epsfig,amssymb,amsfonts, amstext, verbatim,amsopn,multirow,multicol,lipsum}
\usepackage{cite}
\usepackage[font=scriptsize]{caption}
\usepackage{subcaption}
\usepackage{balance}
\usepackage{url}
\usepackage{bbding}
\usepackage{amsfonts}
\usepackage{epsfig}
\usepackage{epstopdf}
\usepackage{setspace}
\usepackage{stmaryrd}
\usepackage{psfrag}	
\usepackage{multirow}
\usepackage{float}
\usepackage{cases}  
\usepackage[process=auto]{pstool}
\usepackage{etoolbox}
\usepackage{algorithm}
\usepackage{algorithmic}
\usepackage{hyperref}
\usepackage{wasysym}[nointegrals]
\usepackage{pifont}
\allowdisplaybreaks
\usepackage{graphicx}
\usepackage{wrapfig}
\usepackage{lscape}
\usepackage{rotating}
\usepackage{epstopdf}

\newtheorem{theo}{Theorem}
\newtheorem{lem}{Lemma}

\newtheorem{prop}{Proposition}
\newtheorem{corol}{Corollary}

\newcommand{\pushright}[1]{\ifmeasuring@#1\else\omit\hfill$\displaystyle#1$\fi\ignorespaces}
\newcommand{\pushleft}[1]{\ifmeasuring@#1\else\omit$\displaystyle#1$\hfill\fi\ignorespaces}
\newtoggle{Conf}

\togglefalse{Conf}

\iftoggle{Conf}{%
}{%
}

\EndPreamble

\usepackage{rotating}

\begin{document}
\title{  Power Scaling Law for  Optical IRSs and Comparison with Optical Relays 
\vspace{-0.3cm}}
\author{Hedieh Ajam\IEEEauthorrefmark{1}, Marzieh Najafi\IEEEauthorrefmark{1}, Vahid Jamali\IEEEauthorrefmark{2},   and Robert Schober\IEEEauthorrefmark{1}\\  
		\IEEEauthorrefmark{1}Friedrich-Alexander-Universi\"at  Erlangen-N\"urnberg, \IEEEauthorrefmark{2}Technical University of Darmstadt
		\vspace{-0.4cm}}
	\maketitle
\begin{abstract}
	The line-of-sight (LOS) requirement of free-space optical (FSO) systems can be relaxed by employing optical relays and optical intelligent reflecting surfaces (IRSs). 	 Unlike  radio frequency (RF) IRSs, which typically exhibit a quadratic power scaling law,  the   power  reflected from FSO IRSs and collected at the receiver lens  may scale quadratically or  linearly  with the IRS size or  may  even saturate at a  constant value. We analyze the power scaling law for optical IRSs and unveil its dependence  on the wavelength,  transmitter (Tx)-to-IRS and  IRS-to-receiver (Rx) distances,  beam waist, and  lens size. 	We compare optical IRSs in different power scaling regimes with  optical relays  in terms of the outage probability,  diversity and  coding gains, and  optimal placement. Our results show that, at the expense of a higher hardware complexity, relay-assisted FSO links yield a better outage performance at  high signal-to-noise-ratios (SNRs), but  optical IRSs can achieve a higher  performance at low SNRs. Moreover, while it  is optimal to place relays equidistant from Tx and Rx, the optimal location of  IRSs  depends on the power scaling regime they operate in.
\end{abstract}
	\section{Introduction}
Due to their  directional narrow laser beams and easy-to-install transceivers, free space optical (FSO) systems   are promising candidates for  high data rate applications, such as wireless front- and back-hauling,  in  next generation  wireless communication networks and beyond \cite{6G}.   FSO systems  require a line-of-sight (LOS) connection between transmitter (Tx) and receiver (Rx)  which can be relaxed by using  optical relays \cite{Safari_relay} or optical intelligent reflecting surfaces (IRSs) \cite{Marzieh_IRS,Marzieh_IRS_jou, AmplifyingRIS_Herald}.  Optical relays process the incident signal and forward an amplified signal to the receiver. For high data rate FSO systems, relays may require high-speed decoding and encoding hardware and/or analog gain units, additional synchronization, and clock recovery \cite{All-optical}.  On the other hand, optical IRSs are planar structures comprised of passive subwavelength elements, known as unit cells, which  can manipulate the  properties of an incident wave such as its phase and polarization \cite{ Marco_survey,IRS-FSO-Herald}. In particular, to redirect an incident beam in a desired direction, the IRS can  apply  a phase shift to the incident wave and adjust the accumulated phase of the reflected wave \cite{TCOM_IRSFSO}. 

For  radio frequency (RF) IRSs,  the received power typically scales quadratically with the IRS area, $\Sigma_\text{irs}$, \cite{RuiZhang_PowerScale,Emil-PowerScale}.  However, for optical IRSs, depending on the receiver lens size,  the location of Tx and Rx with respect to (w.r.t.) the IRS, and the beam waist,  the  received power may scale quadratically ($\mathcal{O}(\Sigma_\text{irs}^2)$) or linearly ($\mathcal{O}(\Sigma_\text{irs})$) with the  IRS size or it may even saturate to a constant value ($\mathcal{O}(1)$) \cite{Vahid_Magazine}. In this  paper, we analyze the power scaling law for optical IRSs in detail  in terms of the system parameters.  

Furthermore, we compare  the performance of relay- and IRS-assisted FSO systems. Such comparisons were made for RF IRSs with  decode-and-forward (DF) and amplify-and-forward (AF) relays in \cite{IRSvsRel_Larsson} and \cite{IRSvsAF_Debbah}, respectively.  However, RF links are fundamentally different from FSO links  in the following aspects: 1) While   spherical/planar RF waves  lead to a uniform power distribution across the IRS, FSO systems employ  Gaussian laser beams, which have a curved wavefront and a non-uniform power distribution \cite{TCOM_IRSFSO}; 2) Unlike in RF links, the variance of the fading affecting FSO links is distance-dependent \cite{Safari_relay}; 3) To reduce hardware complexity,  mostly half-duplex relays are  employed in RF systems,  whereas FSO relays are typically full-duplex \cite{Sahar_Placement}; 4) The electrical size of the IRS (IRS length divided by the wavelength) at optical frequencies is much larger than at RF. 

In this paper, we consider relay- and IRS-assisted FSO systems and  our contributions are summarized as follows: First, we analyze the power scaling law for different IRS sizes, and then, we compare the performance of   IRS- and  relay-assisted  FSO systems  in terms of outage probability. Our results  show that, at the expense of higher hardware complexity,  relay-assisted FSO links yield a higher  diversity gain as the variance of the corresponding distance-dependent fading is smaller compared to that of  IRS-assisted FSO links.  Moreover, the coding gain in IRS-based FSO links may increase with the IRS size depending on the power scaling regime the IRS operates in. We also analyze the  optimal positions of the FSO relays and IRSs for maximization of  the end-to-end  performance. We show that while  relays are optimally  positioned  equidistant from   Tx and Rx \cite{Safari_Placement, Sahar_Placement}, the optimal position of the IRS depends on the  power scaling regime. In particular,  the optimal placement of the IRS  is close to Tx or Rx, close to Tx, and  equidistant  from Tx and Rx if the IRS operates in the quadratic, linear, and saturation power scaling regime, respectively.
\vspace*{-2mm}
\section{System and Channel Models}\label{Sec_System}
	\begin{figure}
		\centering
		\includegraphics[width=0.6\textwidth]{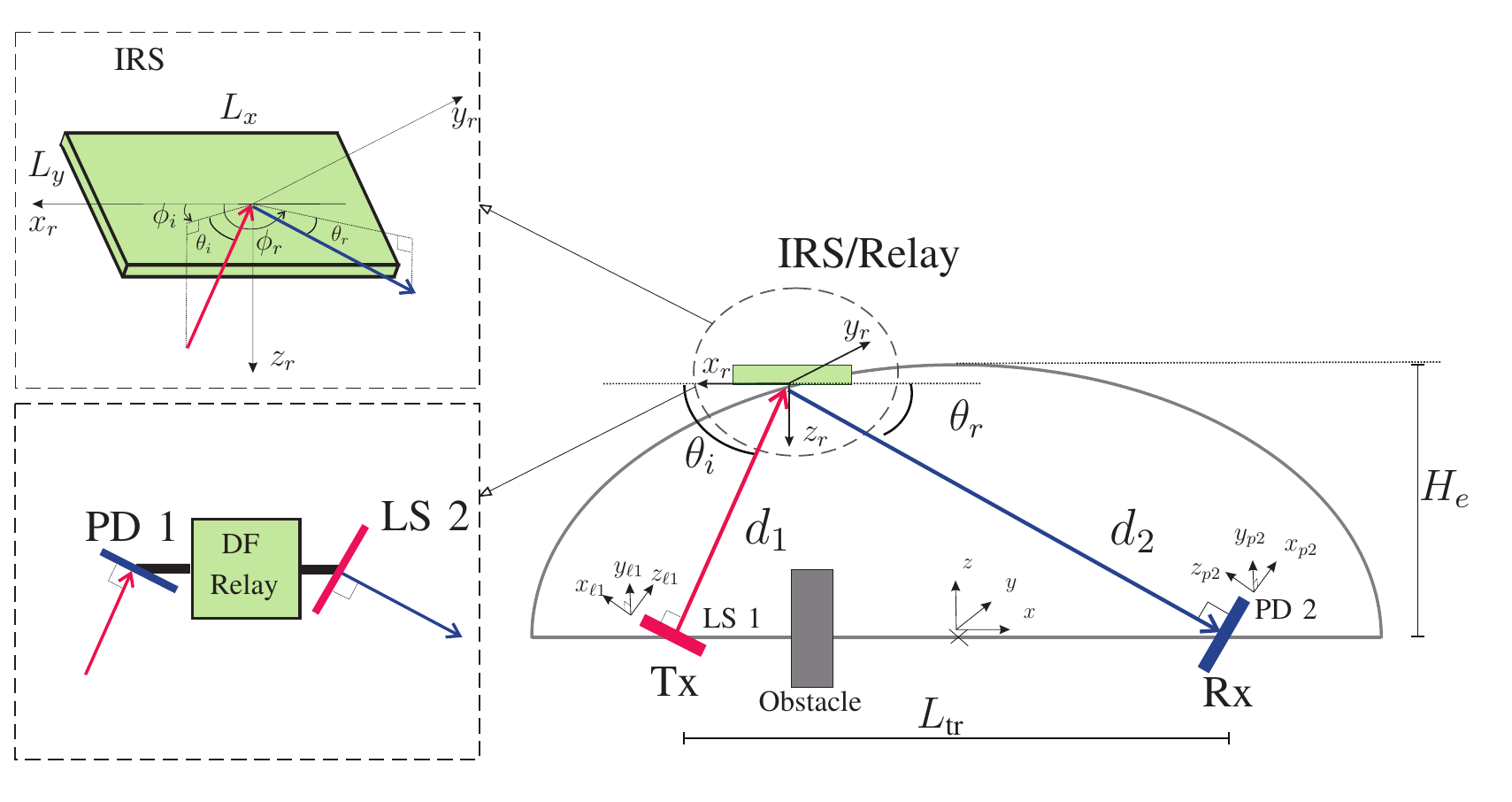}
	\caption{IRS- and relay-based FSO systems with end-to-end distance $d_3=d_1+d_2$ and Tx-to-Rx distance $L_\text{tr}$.}
	\label{Fig:System}\vspace{-0.3cm}
\end{figure} 

We consider two FSO systems, where Tx and  Rx are connected via an IRS and a relay, respectively.  Tx and  Rx are located  on the $x$-axis with  distance  $L_\text{tr}\over 2$ from the origin of the $xyz$-coordinate system, see Fig.~\ref{Fig:System}. Moreover, the centers of the IRS and the relay are located at the origin of the $x_ry_rz_r$-coordinate system, where the $x_ry_r$-plane  is parallel to the $xy$-plane and the $z_r$-axis points in the opposite direction  of the $z$-axis, see Fig.~\ref{Fig:System}.  The Tx is equipped with laser source (LS) 1 emitting a Gaussian laser beam. The beam axis  intersects with the $x_ry_r$-plane at distance  $d_1$ and in direction $\mathbf{\Psi}_{i}=\left(\theta_{i},\phi_{i}\right)$, where $\theta_i$ is the angle between the $x_ry_r$-plane and the beam axis, and $\phi_{i}$ is the angle between the projection of the beam axis on the $x_ry_r$-plane and the  $x_r$-axis. Moreover, the Rx is equipped with photo-detector (PD) 2 and  a circular  lens of radius $a$.  The lens  of the Rx   is located at distance $d_{2}$ from the origin of the $x_ry_rz_r$-coordinate system. The normal vector of  the lens plane points in direction $\mathbf{\Psi}_{r}=\left(\theta_{r}, \phi_{r}\right)$, where $\theta_{r}$ is the angle between the $x_ry_r$-plane and the normal vector, and $\phi_{r}$ is the angle between the projection of the normal vector on the $x_ry_r$-plane and the $x_r$-axis. We assume that the  Rx lens plane is always perpendicular to the  axis of the received beam.  In the following, we describe the  relay- and IRS-assisted FSO systems more in detail.
\subsection{Relay-Assisted FSO System Model}
We assume  the Tx is connected to the Rx via a full-duplex DF relay\footnote{In this work, we assume  DF relaying  which may yield better or almost  similar performance  as AF relaying depending on the channel conditions \cite{Sahar_Placement}.} where the relay receives the transmitted symbols via PD 1, re-encodes the signal, and transmit it via LS 2   to  PD 2  at the Rx.  PD 1  is equipped with a circular lens of radius $a$ which is always perpendicular to the axis of the received beam. 
Assuming an intensity modulation and direct detection (IM/DD) system,  the received signal intensity  at the relay, $y_{1}$, is given by 
\begin{IEEEeqnarray}{rll}
y_{1}&=\sqrt{P_1}h_1 s_1+n_{1},
	\label{system_eq_SR}
\end{IEEEeqnarray}
where $s_1$ is  the  symbol transmitted by LS 1   with $\mathbb{E}\{|s_1|^2\}=1$, $P_1$ is the transmit power of LS 1, $h_1\in \mathcal{R}^+$  denotes the  gain of the Tx-to-relay link,  and $n_1\sim \mathcal{N} \left(0, \sigma_n^2\right)$ is the additive white Gaussian noise (AWGN) at PD 1 with zero mean and variance $\sigma_n^2$. Here, $\mathbb{E}\{\cdot\}$ denotes expectation. Then, the received signal intensity  at  PD 2, $y_{2}$, is given by
\begin{IEEEeqnarray}{rll}
y_{2}&=\sqrt{P_2}h_2 s_2+n_2,
	\label{system_eq_RD}
\end{IEEEeqnarray}
where  $s_2$ is the  signal transmitted by the relay with $\mathbb{E}\{|s_2|^2\}=1$, $h_{2}\in \mathcal{R}^+$ denotes the FSO channel gain of the relay-to-Rx link,  $P_2$ is the transmit power of LS 2, and $n_2\sim \mathcal{N} \left(0, \sigma_n^2\right)$ is the AWGN noise at PD 2. $P_1$ and $P_2$ are chosen such that   $P_1=P_2=\frac{P_\text{tot}}{2}$, where $P_\text{tot}$ is the total transmit power. 
\subsection{IRS-Assisted FSO System Model}
We assume the Tx  is connected via a single IRS to the Rx. The size of the IRS is $\Sigma_\text{irs}=L_{x}\times L_{y}$,   where $L_x$ and $L_y$ are the length of the IRS in  $x_r$- and $y_r$-direction, respectively. The IRS is comprised of  passive elements and assuming $L_x,L_y\gg  \lambda$, the IRS can be modeled as a continuous surface with  continuous linear phase shift profile.  To realize anomalous reflection, we focus on a linear phase shift design denoted by $\Phi_\text{irs}({\mathbf{r}_r})=k\left(\Phi_x x_r+\Phi_y y_r+\Phi_0\right)$, where $\mathbf{r}_r=(x_r,y_r,0)$ denotes a point in the $x_ry_r$-plane. To redirect the beam from  Tx  direction $\mathbf{\Psi}_i$ to Rx direction $\mathbf{\Psi}_r$, the phase gradients are chosen as  $\Phi_x=\cos(\theta_{i})\cos(\phi_{i})+\cos(\theta_r)\cos(\phi_r)$, $\Phi_y=\cos(\theta_{i})\sin(\phi_i)+\cos(\theta_r)\sin(\phi_r)$, and the constant is $\Phi_0=d_1+d_2$, see \cite{TCOM_IRSFSO}. 
Assuming an IM/DD FSO system,   the  received signal intensity $y_{3}$ at  PD 2 for the IRS-assisted FSO link is given by
\begin{IEEEeqnarray}{rll}
	y_{3}=\sqrt{P_\text{tot}}h_3 s_1 +n_2,
	\label{system_eq_IRS}
\end{IEEEeqnarray}
where   $h_3 \in \mathcal{R}^+$ is the end-to-end channel gain between Tx, IRS, and Rx  and  LS 1 transmits with  power $P_\text{tot}$.
\subsection{Channel Model}
FSO channels are impaired by  geometric and misalignment losses (GML),  atmospheric loss, and  atmospheric turbulence induced fading \cite{IRS_FSO_WCNC}.   Thus, the point-to-point FSO channel gains  are modeled as follows
\begin{IEEEeqnarray}{rll}
	h_i=\zeta h_{p,i} h_{a,i} h_{\text{gml},i}, \quad i\in\{1,2,3\},
	\label{ch}
\end{IEEEeqnarray}
where $\zeta$ is the PD responsivity, $h_{a,i}$ represents the random atmospheric turbulence induced fading,  $h_{p,i}$ is the atmospheric loss, and  $h_{\text{gml},i}$ characterizes the GML. 
\subsubsection{Atmospheric Loss}
The atmospheric loss characterizes the laser beam energy loss due to absorption and scattering and  is given by	$h_{p,i}=10^{-\frac{\kappa d_i}{10}},  i\in\{1,2,3\},$ where $\kappa$ is the attenuation coefficient and   $d_{3}=d_1+d_2$ denotes the end-to-end link distance.
\subsubsection{Atmospheric Turbulence}
The variations of the refractive index along the propagation path due to changes in temperature and pressure cause atmospheric turbulence which is analogous to  the fading in RF systems. Assuming $h_{a,i}$ is  a Gamma-Gamma distributed random variable, its cumulative distribution function  (CDF)  is given by \cite{Alouini-GG}
\begin{IEEEeqnarray}{rll}
	F_{h_{a,i}}(x)=\frac{1}{\Gamma(\alpha_i)\Gamma(\beta_i)} G_{1,3}^{2,1}\left(\alpha_i\beta_i{{x}}\left|\begin{array}{@{}l@{}} \ \ \ \ \ \ \ 1 \\ \alpha_i,\beta_i,0 \\ \end{array} \right. \right),
	\label{EQ:CDF_GG}
\end{IEEEeqnarray}
where $\Gamma(\cdot)$ denotes the Gamma function and  $G(\cdot|\cdot)$ is the Meijer G-function  \cite{integral}.  Here,  the small and large scale turbulence parameters $\alpha_i$ and $\beta_i$  depend on the Rytov variance $\sigma_{R,i}^2=1.23C_n^2k^{7\over 6}d_i^{11\over 6}$, where $k=\frac{2\pi}{\lambda}$ is the wave number, $\lambda$ is the wavelength, and $C_n^2$ is the refractive-index structure constant  \cite{Sahar_Placement}. 
\vspace*{-2mm}
\subsubsection{GML}
The GML coefficient comprises the deterministic geometric loss due to the divergence of the laser beam along the transmission path and the random misalignment loss due to transceiver sway \cite{Vahid_Magazine}. Here,  we ignore the misalignment  loss 
and determine the geometric loss for the relay- and IRS-based links in the following.

Assuming the waist of the Gaussian beam is larger than the wavelength, $w_{oi}\gg \lambda$, the electric field of the Gaussian laser beam  emitted by the  $i$-th LS, $\forall i\in\{1, 2\}$,  is given by \cite{Goodmanbook}
\begin{IEEEeqnarray}{rll}
E_{\ell i}\!\left(\mathbf{r}_{\ell i}\right)&= \frac{4\eta P_\text{tot}}{n\pi{w}(z_{\ell i})} e^{-{x_{\ell i }^2+y_{\ell i}^2\over w^2(z_{\ell i })}-j	k\left(\!\!z_{\ell i}+{x_{\ell i}^2+y_{\ell i }^2\over 2R(z_{\ell i })}\right)-\tan^{-1}\left(\frac{z_{\ell i}}{z_{R i}}\right)},\qquad
	\label{Gauss}
\end{IEEEeqnarray}
where $\eta$ is the free-space impedance, $n=1$ and  $2$ for the IRS- and relay-assisted links, respectively, and $\mathbf{r}_{\ell i}=(x_{\ell i }, y_{\ell i }, z_{\ell i})$ is a point in a  coordinate system, which has its origin at the $i$-th LS.  The $z_{\ell i }$-axis of this coordinate system  is along the beam axis, the $y_{\ell i }$-axis is parallel to the intersection line of the $i$-th LS plane and the $x_ry_r$-plane, and the $x_{\ell i}$-axis  is orthogonal to the $y_{\ell i}$- and $z_{\ell i}$-axes.   $w(z_{\ell i})=w_{o i}\Big[{1+\left(\frac{z_{\ell i}}{z_{R i}}\right)^2}\Big]^{1/2}$ is the beamwidth at distance $z_{\ell i}$, $R(z_{\ell i})=z_{\ell i}\Big[1+\left(\frac{z_{R i}}{z_{\ell i}}\right)^2\Big]$ is the radius of the curvature of the beam's wavefront, and  $z_{R i}=\frac{\pi w_{o i}^2}{\lambda}$ is the Rayleigh range.

Assuming the lenses at the relay and the Rx are always perpendicular to the incident beam axes, respectively, the GML coefficients  of the  Tx-to-relay link, $h_{\text{gml},1}$, and the relay-to-Rx link, $h_{\text{gml},2}$, are given by \cite{Farid_fso}
\begin{IEEEeqnarray}{rll}
	h_{\text{gml},i}=\left[\text{erf}\left(\sqrt{\frac{{\pi}}{2}}\frac{a}{w(d_i)}\right)\right]^2,  \,\quad i\in\{1,2\},
	\label{gain_rel}
\end{IEEEeqnarray}
where  $\text{erf}(\cdot)$ denotes the error function \cite{integral}. Moreover, the GML factor of the IRS-assisted FSO link is given by
\begin{IEEEeqnarray}{rll}
	h_{\text{gml},3}={1 \over 2 \eta P_\text{tot}}\iint\nolimits_{\mathcal{A}_{p2}}\lvert E_{r}\left(\mathbf{r}_{p2}\right) \rvert^2 \, \mathrm{d}\mathcal{A}_{p2}, \quad
	\label{gain}
\end{IEEEeqnarray}
 where $\mathcal{A}_{p2}$  denotes the area of the lens of  PD 2 and $\mathbf{r}_{p2}=(x_{p2}, y_{p2}, z_{p2})$ denotes a   point on the Rx lens plane.  The origin of the $x_{p2}y_{p2}z_{p2}$-coordinate system is  the center of the  Rx lens and the $z_{p2}$-axis  is parallel to  the normal vector of the Rx lens plane. We assume that the $y_{p2}$-axis is parallel to the intersection line of the lens plane and the $x_ry_r$-plane and the $x_{p2}$-axis is perpendicular to the $y_{p2}$- and $z_{p2}$-axes.
 $E_{r}\left(\cdot\right)$ is  the electric field of the  beam reflected by the IRS and received at the lens of the Rx  and  is given by  \cite{TCOM_IRSFSO}
 \iftoggle{Conf}{%
 	\begin{IEEEeqnarray}{rll}
 		E_{r}&\left(\mathbf{r}_{p2}\right)=\!\!C_r\!\!\!\iint\limits_{\mathbf{r}_r\in\Sigma_\text{irs}}\!\!\!\!\! E_{\text{in}}({\mathbf{r}_r})e^{-j\Phi_\text{irs}(\mathbf{r}_r)-jk d_2 \left[1+\frac{|\mathbf{r}_{p2}-\mathbf{r}_r|^2}{d_2^2}\right]^{1/2}} \!\!\!\!\! \mathrm{d}{x}_r\mathrm{d}{y}_r, \quad
 		\label{Huygens}
 		\vspace{-2mm}
 	\end{IEEEeqnarray}
 }{
\begin{IEEEeqnarray}{rll}
	E_{r}\left(\mathbf{r}_{p2}\right)=C_r\iint_{({x}_r,{y}_r)\in\Sigma_\text{irs}} E_{\text{in}}({\mathbf{r}_r}) \exp\left(-jk d_2 \left[1+\frac{|\mathbf{r}_{p2}-\mathbf{r}_r|^2}{d_2^2}\right]^{1/2} \right) e^{-j\Phi_\text{irs}(\mathbf{r}_r)} \mathrm{d}{x}_r\mathrm{d}{y}_r, \quad
	\label{Huygens}
\end{IEEEeqnarray}}
where $C_r={\sqrt{\sin(\theta_r)}}/({j\lambda d_2})$, $E_{\text{in}}({\mathbf{r}_r})$ is the  incident electric field  on the IRS given by \cite[Eq.~(6)]{TCOM_IRSFSO}.  A closed-form solution for (\ref{gain}) is given by \cite[Eq.~(21)]{TCOM_IRSFSO}.  In this paper, to gain insight for FSO system design and to determine the corresponding power scaling law, we analyze (\ref{Huygens}) for different IRS sizes, $\Sigma_\text{irs}$, and lens sizes, $\Sigma_\text{lens}$: 1) $\Sigma_\text{irs}\ll A_\text{in}$ and $\Sigma_\text{lens}\ll A_\text{rx}$, 2)  $\Sigma_\text{irs}\ll A_\text{in}$ and $\Sigma_\text{lens}\gg A_\text{rx}$, 3) $\Sigma_\text{irs}\gg A_\text{in}$, where  $A_\text{in}=\pi w_{\text{in},x}w_{\text{in},y}$ and $A_\text{rx}=\pi w_{\text{rx},x}w_{\text{rx},y}$ are the areas of the beam footprint on the IRS and  lens, respectively. Here, $w_{\text{in},x}=\frac{w(d_1)}{\sin(\theta_i)}$ and $w_{\text{in},y}=w(d_1)$ are the incident beam widths on the IRS in $x$- and $y$-direction, respectively. Moreover, $w_{\text{rx},x}$ and $w_{\text{rx},y}$ are the received beam widths at the lens in $x_p$- and $y_p$-direction, respectively.
\vspace*{-2mm}
\section{ Power Scaling Laws for  Optical IRSs}\label{Sec:Power Regimes}
In this section, we analyze the received power  and  show that the GML and the received power at the lens may scale quadratically or linearly with the IRS size or may remain constant.
\subsection{Quadratic Power Scaling Regime}
We first consider the case when the  IRS  is small,   and    hence, only a small fraction of the Gaussian beam is received at the IRS. The following lemma  provides an  approximation  for the GML.
\begin{lem}\label{Lemma1}
	Assuming  $L_x\ll w_{\text{in},x}$, $L_y\ll w_{\text{in},y}$, and $\Sigma_\text{lens}\ll A_\text{rx}$ 
	,	the  GML for the IRS-assisted  link, $h_{\text{gml},3}$,  can be approximated by $\tilde{G}_1$, which is given as follows
	 \iftoggle{Conf}{%
	 		\begin{IEEEeqnarray}{rll}
	 		\tilde{G}_1&=C_1\left[c_1a \sqrt{\pi} \text{Si}\left(c_1a\sqrt{\pi}\right)+\cos\left(c_1a\sqrt{\pi} \right)-1\right]\nonumber\\
	 		&\times\left[c_2a \sqrt{\pi} \text{Si}\left(c_2a\sqrt{\pi}\right)+\cos\left(c_2a\sqrt{\pi}\right)-1\right],
	 		\label{EQ:Lem1}
	 	\end{IEEEeqnarray}
	 }{
	\begin{IEEEeqnarray}{rll}
		\tilde{G}_1=&C_1\times\left[c_1a \sqrt{\pi} \text{Si}\left(c_1a\sqrt{\pi}\right)+\cos\left(c_1a\sqrt{\pi} \right)-1\right]\nonumber\\
		&\times\left[c_2a \sqrt{\pi} \text{Si}\left(c_2a\sqrt{\pi}\right)+\cos\left(c_2a\sqrt{\pi}\right)-1\right],
		\label{EQ:Lem1}
	\end{IEEEeqnarray}
}
	where $C_1=\frac{16 d_2^2\tilde{G}_2}{\pi^3 a^2 k^2 L_xL_y \left|\sin(\theta_r)\right|}$, $c_1=\frac{k\sin(\theta_r)L_x}{2d_2}$, $c_2=\frac{kL_y}{2d_2}$,  $\text{Si}(\varkappa)=\int_{0}^{\varkappa}\frac{\sin(t)}{t} \mathrm{d}t$ denotes the sine integral function, and $\tilde{G}_2=\text{erf}\left(\frac{\sqrt{2}}{2}\frac{L_x\sin(\theta_i)}{w(d_1)}\right)\text{erf}\left(\frac{\sqrt{2}}{2}\frac{L_y}{w(d_1)}\right)$.
\end{lem}  
\begin{IEEEproof}
	The proof is given in Appendix \ref{App1}.
\end{IEEEproof}

 In this case, due to the small IRS size, the amplitude of the received electric field is the product of two sinc-functions, see (\ref{EQ:Proof_Lem1_sinc}) in Appendix \ref{App1}.  Thus,  the coherent superposition of the signals reflected from all points on the IRS at the lens introduces a beamforming gain.  By increasing the IRS size,   the beamwidth of the sinc-shaped beam at the lens decreases, which in turn  increases the peak amplitude of the beam causing the beamforming gain. In addition to this beamforming gain, a larger IRS surface collects more power from the incident beam which results in a quadratic  scaling of the received power  with the IRS size. This behavior is analytically confirmed in the following corollary. 
\begin{corol}\label{Corollary1}
	For $L_x,L_y\to 0$ and $d_1\gg z_{R1}$, $\tilde{G}_1$ can be approximated by
	\begin{IEEEeqnarray}{rll}
	{G}_1&= 4\pi\times\frac{4\pi \Sigma_\text{irs}^2 \left|\sin(\theta_r)\right|\left|\sin(\theta_i)\right| }{\lambda^4} \times g_{\text{LS}} \times g_\text{PD}, \quad
		\label{EQ:ApproxG_1}
	\end{IEEEeqnarray}
where $g_\text{LS}=\frac{2\pi w_o^2}{4\pi d_1^2}$ and $g_\text{PD}=\frac{\pi a^2}{4\pi d_2^2}$. Since, $G_1$ scales quadratically with the  IRS size $\Sigma_\text{irs}$, we refer to this regime as the ``quadratic power scaling regime''.
\end{corol}
\begin{IEEEproof}
	 We  substitute in (\ref{EQ:Lem1}) the Taylor series expansions of $\text{Si}(x)\approx x$ and $\cos(x)\approx 1-\frac{x^2}{2}$  and use the Taylor series expansion of $\text{erf}(x)\approx \frac{2}{\sqrt{\pi}} x$. Then, assuming $d_1\gg z_{R1}$, we can substitute  $w(d_1)\approx \frac{d_1 \lambda}{\pi w_{o1}}$ and this completes the proof.
\end{IEEEproof}  
The quadratic scaling law shown above is in agreement with the power scaling laws shown in  \cite[Eq.~(2), (10)]{Marzieh_Poor} and \cite[Eq.~(48)]{Emil-PowerScale} for RF IRSs.
\subsection{Linear Power Scaling Regime}\label{Sec:G2}
As the size of the IRS increases,  the beamforming gain cannot further increase the received power and the larger IRS size only collects the rest of the incident power on the IRS.  In this regime, the lens is much larger than the beam footprint at the Rx such that the total  power incident on the IRS is received at the Rx lens.  In the following lemma, we determine the GML  for this case.

\begin{lem}\label{Lemma2}
	Assuming  $L_x\ll w_{\text{in},x}$, $L_y\ll w_{\text{in},y}$ and $\Sigma_\text{lens}\gg A_\text{rx}$, the GML factor $h_{\text{gml},3}$, is approximated by $\tilde{G}_2$ and  given by
	\begin{IEEEeqnarray}{rll}
		\tilde{G}_2=\text{erf}\left(\frac{\sqrt{2}}{2}\frac{L_x\sin(\theta_i)}{w(d_1)}\right)\text{erf}\left(\frac{\sqrt{2}}{2}\frac{L_y}{w(d_1)}\right).\quad
		\label{EQ:Lem2}
	\end{IEEEeqnarray}
\end{lem} 
\begin{IEEEproof}
	The proof is provided in Appendix \ref{App2}.
\end{IEEEproof}
In (\ref{EQ:Lem2}), $\tilde{G}_2$  is  the normalized incident power on the IRS which  provides a tight upper bound on the GML, $h_{\text{gml},3}$,  in the considered case. To determine the slope of this function  w.r.t. the IRS size, we approximate (\ref{EQ:Lem2}) in the following corollary.
\begin{corol}\label{Corollary2}
	Assuming $\frac{L_x}{w(d_1)}, \frac{L_y}{w(d_1)} \to 0$, $\tilde{G}_2$  can be approximated as follows
	\begin{IEEEeqnarray}{rll}
	{G}_2=  \frac{4\pi \Sigma_\text{irs} \left|\sin(\theta_i)\right|}{\lambda^2}\times g_\text{LS}.\quad
		\label{EQ:ApproxG_2}
	\end{IEEEeqnarray}
Since $G_2$ scales linearly with the IRS size, $\Sigma_\text{irs}$,  we refer to this regime as the ``linear power scaling regime''.
\end{corol}
\begin{IEEEproof}
	We apply the Taylor series expansion of $\lim\limits_{x\to 0}\text{erf}(x)\approx \frac{2}{\sqrt{\pi}} x$ in (\ref{EQ:Lem2}). This completes the proof.
\end{IEEEproof} 
\subsection{Saturated Power Scaling Regime}
For the case, when the IRS size is very large, such that the lens size is the limiting factor for the received power,  the GML  is given in  the following lemma.
\begin{lem}\label{Lemma3}
	Assuming  $L_x\gg w_{\text{in},x}$ and $L_y\gg w_{\text{in},y}$, the GML, $h_{\text{gml},3}$,  is approximated by $G_3$, which is given by
	\begin{IEEEeqnarray}{rll}
		G_3=\text{erf}\left(\sqrt{\frac{\pi}{2}}\frac{a}{W_{\text{eq},x}}\right)\text{erf}\left(\sqrt{\frac{\pi}{2}}\frac{a}{W_{\text{eq},y}}\right),
		\label{EQ:Lem3}
	\end{IEEEeqnarray}
 \iftoggle{Conf}{%
 	where $\!\!W_{\text{eq},x}\!\!\!\!=\!\!\!\!\!\!\frac{w(d_1)\left|\sin(\theta_r)\right|}{\left|\sin(\theta_i)\right|} \!\left[\left(\frac{\Lambda_1 \sin^2(\theta_{i})}{ \sin^2(\theta_{r})}\right)^2\!\!\!\!\!+\!\!\left(\frac{\Lambda_2 \sin^2(\theta_{i})}{ \sin^2(\theta_{r})}+1\right)^2\right]^{1\over2}$,\,\, $W_{\text{eq},y}\!\!\!=\!\!w(d_1) \!\left[\Lambda_1^2+\left(\Lambda_2 +1\right)^2\right]^{1\over 2}\!\!\!\!$, $\Lambda_1=\frac{2d_2}{kw^2(d_1)}$, and $\Lambda_2=\frac{d_2}{R(d_1)}$.
 }{
 where $W_{\text{eq},x}=\frac{w(d_1)\left|\sin(\theta_r)\right|}{\left|\sin(\theta_i)\right|} \left[\left(\frac{\Lambda_1 \sin^2(\theta_{i})}{ \sin^2(\theta_{r})}\right)^2+\left(\frac{\Lambda_2 \sin^2(\theta_{i})}{ \sin^2(\theta_{r})}+1\right)^2\right]^{1/2}\!\!\!\!\!\!$, $W_{\text{eq},y}=w(d_1) \left[\Lambda_1^2+\left(\Lambda_2 +1\right)^2\right]^{1/2}\!\!\!\!$, $\Lambda_1=\frac{2d_2}{kw^2(d_1)}$, and $\Lambda_2=\frac{d_2}{R(d_1)}$.}
\end{lem} 
\begin{IEEEproof}
	The proof is provided in Appendix \ref{App3}.
\end{IEEEproof}
The above lemma shows that, in the considered case, the normalized received power at the lens does not depend on the IRS size. Thus, we refer to this regime as the ``saturation power scaling regime''.
\vspace*{-2mm}
\subsection{ GML Coefficient of IRS-Assisted FSO Link ($h_{\text{gml},3}$)}
In the following preposition, we analyze the IRS sizes for which the quadratic, linear, and saturation power scaling laws are valid.  
\begin{prop}\label{Theorem1}
	If  $G_3\geq \frac{2d_2^2w_{o1}^2|\sin(\theta_i)|}{d_1^2a^2|\sin(\theta_r)|}$,  $h_{\text{gml},3}$  scales   with the IRS size,  $\Sigma_\text{irs}$,  as follows
	\begin{IEEEeqnarray}{rll}
		h_{\text{gml},3}\approx
		\begin{cases}
			{G}_1 \, \text{in} \, (\ref{EQ:ApproxG_1}),\quad &  \Sigma_\text{irs}< S_1,\\
			{G}_2 \, \text{in} \, (\ref{EQ:ApproxG_2}),\quad &  S_1\leq \Sigma_\text{irs}\leq S_2, \\
			G_3 \, \text{in} \, (\ref{EQ:Lem3}), \quad & \Sigma_\text{irs} > S_2,
		\end{cases}		
		\label{EQ:Theo1}
	\end{IEEEeqnarray}
where $S_1=\frac{\lambda^2 d_2^2}{\pi a^2\left|\sin(\theta_r)\right|}$ and $S_2= \frac{\pi G_3 w^2(d_1)}{2\left|\sin(\theta_{i})\right|}$ are the boundary IRS sizes where the  transition from  quadratic to  linear and from  linear to saturation power scaling  occurs, respectively. If $G_3< \frac{2d_2^2w_{o1}^2|\sin(\theta_i)|}{d_1^2a^2|\sin(\theta_r)|}$, the GML scales only quadratically  with the IRS size,  $\Sigma_\text{irs}$, or is a constant as follows
\begin{IEEEeqnarray}{rll}
	h_{\text{gml},3}\approx
	\begin{cases}
	{G}_1 \, \text{in} \, (\ref{EQ:ApproxG_1}),\quad & \Sigma_\text{irs}\leq S_3,\\
		G_3 \, \text{in} \, (\ref{EQ:Lem3}), \quad & \Sigma_\text{irs}> S_3,
	\end{cases}		
	\label{EQ:Theo1b}
\end{IEEEeqnarray}
where   $S_3=\frac{\sqrt{G_3}\lambda d_2 w(d_1)}{a \sqrt{2\sin(\theta_i)\sin(\theta_r)}}$ is the IRS size for which the transition from  quadratic to linear power scaling  occurs.
\end{prop} 
\begin{IEEEproof}
	The boundaries $S_1$ and $S_2$  are derived as the intersection points of (\ref{EQ:ApproxG_1}),  (\ref{EQ:ApproxG_2}) and (\ref{EQ:ApproxG_2}), (\ref{EQ:Lem3}), respectively. If $S_2<S_1$,  linear power scaling does not occur. $S_3$ is the intersection point of (\ref{EQ:ApproxG_1})  and (\ref{EQ:Lem3}). This leads to (\ref{EQ:Theo1b}) and completes the proof.
\end{IEEEproof}
The above preposition  shows  how the received power scales with the IRS size for given system parameters such as the LS parameters, $w_{o1}$ and $\lambda$, the lens radius, $a$,  the  distances, $d_1$ and $d_2$, and the angles $\theta_i$ and $\theta_r$. Moreover, due to the large electrical size of the lens ($\frac{\pi a^2}{\lambda^2}\approx 10^{8}$) in FSO systems, the boundary IRS size, $S_1$, is comparatively small, and thus, optical IRSs of sizes $10$ cm-1 m   typically operate  in the linear  or saturated power scaling regimes.  Unlike FSO systems, the electrical size of RF antennas is comparatively small ($\approx 1$) which leads to large values for $S_1$, and thus, even RF IRSs of having large sizes of $1-10$ m   operate in the quadratic power scaling regime.
 
\vspace*{-2mm}
\section{Diversity and Coding Gains}\label{Sec_Diversity}
For a fixed transmission rate, the outage probability is defined as the probability that the instantaneous SNR, $\gamma$,   is smaller than a threshold SNR, $\gamma_{th}$, i.e., $P_\text{out}=\text{Pr}\left(\gamma<\gamma_{th}\right)$. At high SNR, the outage probability can be approximated as $\lim\limits_{\bar{\gamma}\to\infty}P_\text{out}\approx(C\bar{\gamma})^{-D}$, where $C$ is the coding gain,  $\bar{\gamma}$ is the average transmit SNR, and $D$ is the diversity gain. In the following, we compare the  diversity and coding gains of IRS- and relay-assisted FSO systems. 
\subsection{Outage Performance of IRS-assisted Link }
For the IRS-assisted FSO link in (\ref{system_eq_IRS}), the average received power is $\bar{\gamma}_3=\bar{\gamma} \tilde{\gamma}_3$, where $\bar{\gamma}=\frac{P_\text{tot}}{\sigma_n^2}$ and $\tilde{\gamma}_3=h_{\text{gml},3}^2h_{p,3}^2$, and thus, the outage probability is given by \cite{Sahar_Placement}
\begin{IEEEeqnarray}{rll}
	P_\text{out}^\text{irs}=F_{h_{a,3}}\left(\sqrt{{\gamma_{th}}/{\bar{\gamma}_3}}\right),
	\label{EQ:Outage_IRS}
\end{IEEEeqnarray}
where $F_{h_{a,3}}(\cdot)$ is given in (\ref{EQ:CDF_GG}). Thus, using the same approach as in \cite{Sahar_Placement}, the diversity gain, $D_\text{irs}$, and the coding gain, $C_\text{irs}$, of the IRS-assisted FSO link  respectively can be obtained as
\begin{IEEEeqnarray}{rll}
D_\text{irs}=\frac{\varrho_3}{2},\quad
C_\text{irs}=\frac{\tilde{\gamma}_3}{\gamma_{th}(\tau_3 \varrho_3)^2}\left(\frac{\Gamma\left(\tau_3-\varrho_3\right)}{\Gamma(\tau_3)\Gamma(\varrho_3+1)}\right)^{-1/D_\text{irs}},
\label{EQ:Diversity_IRS}	
\end{IEEEeqnarray}	
where $\varrho_3=\min\{\alpha_3, \beta_3\} $ and $\tau_3=\max\{\alpha_3, \beta_3\}$. 
\subsection{Outage Performance of Relay-assisted Link }
The outage probability of a relay-assisted FSO link is given by \cite{Sahar_Placement}
\begin{IEEEeqnarray}{rll}
	P_\text{out}^\text{rel}=1-\prod_{i=1}^{2}\left(1-F_{h_{a,i}}\left(\sqrt{{\gamma_{th}}/{\bar{\gamma}_i}}\right)\right),
	\label{EQ:Outage_relay}
\end{IEEEeqnarray}
where $\bar{\gamma}_i=\bar{\gamma}\tilde{\gamma}_i$ and $\tilde{\gamma}_i=\frac{1}{2}h_{\text{gml},i}^2h_{p,i}^2$, $\forall i\in\{1,2\}$.
 Moreover,  the coding gain, $C_\text{rel}$, and diversity gain, $D_\text{rel}$,  of the relay-assisted FSO link are given as follows \cite{Sahar_Placement}
\begin{IEEEeqnarray}{rll}
	C_\text{rel}&=\begin{cases}
		C_{\text{rel},\iota}&\!\!\!\varrho_1\neq\varrho_2,\\
		\!\!\left(\!\sum\limits_{i=1}^{2}\left(C_{\text{rel},i}\right)^{-D_\text{rel}}\!\!\right)^{\!\!\!-1\over D_\text{rel}}&\!\!\!\varrho_1=\varrho_2,	
	\end{cases}\!\!, D_\text{rel}&=\min\{\frac{\varrho_1}{2},\frac{\varrho_2}{2}\},\quad\,
	\label{EQ:Diversity_Relay}	
\end{IEEEeqnarray}	
 where $C_{\text{rel},i}=\frac{\tilde{\gamma}_i}{\gamma_{th}}\left(\frac{\Gamma(\tau_i-\varrho_i)\left(\tau_i\varrho_i/\mu_i\right)^{\varrho_i}}{\Gamma(\tau_i)\Gamma(\varrho_i+1)}\right)^{-2/\varrho_i}$,  $\varrho_i=\min\{\alpha_i, \beta_i\} $, $\tau_i=\max\{\alpha_i, \beta_i\}$, $\forall i\in\{1,2\}$, and $\iota=\mathrm{arg} \min\limits_{i\in\{1,2\}}\{\varrho_i\}$. 

For larger distances, the Gamma-Gamma  fading parameters, $\alpha_i$ and $\beta_i$, become smaller, see \cite[Eq.~(33)]{Sahar_Placement}. Thus, $\varrho_3<\varrho_1,\varrho_2$, and as shown in (\ref{EQ:Diversity_IRS}) and (\ref{EQ:Diversity_Relay}), the diversity gain of the relay-assisted link is larger than that of the IRS-assisted link. Thus, a relay-assisted link   outperforms an IRS-assisted  link at high SNRs. However,  depending on the system parameters, the coding gain of the IRS-assisted FSO link  may be larger than that of the relay-assisted link, which can boost the performance at low SNRs.
 
\section{Optimal Operating Position of IRS and Relay}\label{Sec:Placement}
Exploiting the analysis in Sections \ref{Sec:Power Regimes} and \ref{Sec_Diversity}, we determine the optimal  positions of the center of  IRS and relay, denoted by $(x_o^*, z_o^*)$ in the $xyz$-coordinate system, where the outage probability of the IRS- and relay-assisted links  at high SNR is minimized, respectively. For a fair comparison, we assume that the end-to-end distance $d_3$ is constant, i.e., IRS and relay are located on an ellipse, see Fig.~\ref{Fig:System}. Thus, we formulate the following optimization problem
\begin{IEEEeqnarray}{rll}
	&\min_{x_o, z_o} P_\text{out}^i\approx \left(C_i \bar{\gamma}\right)^{-D_i},\quad i\in\{\text{rel}, \text{irs}\},\nonumber\\
	&\text{s.t.}\quad \frac{x_o^2}{d_3^2}+\frac{z_o^2}{d_3^2-L_\text{tr}^2}=\frac{1}{4},
	\label{EQ:Opt_prob}
\end{IEEEeqnarray}
\subsection{Optimal Position of  IRS}
Given that   parameters $\alpha_3$ and $\beta_3$   only depend on the end-to-end distance \cite{IRS-FSO-Herald, TCOM_IRSFSO},  the outage probability of the IRS-based link in (\ref{EQ:Opt_prob}) is minimized if  $h_{\text{gml},3} $ is maximized. Thus, the optimal position of the IRS as a function of its size is given in the following theorem.
\begin{theo}\label{Lemma4}
	The  optimal position of the center of the IRS, $(x_o^*, z_o^*)$,  depends on the size of the IRS and is given by 
	\begin{IEEEeqnarray}{rll}
		(x_o^*,z_o^*)\!\!=\!\!
		\begin{cases}
		 \left(\pm\frac{\sqrt{2\rho_1}d_{3}}{4L_\text{tr}},z^*_1 \right), &\Sigma_\text{irs}\leq S_1 \vee \Sigma_\text{irs}\leq S_3 \\
		 \left(\frac{d_{3}}{8L_\text{tr}}\left(d_{3}- \rho_2\right), z^*_2 \right), 	&S_1\leq \Sigma_\text{irs}\leq S_2,\\
		 \left(0, H_e\right),&S_2\leq \Sigma_\text{irs} \vee S_3\leq\Sigma_\text{irs},
		\end{cases}	\qquad
	\label{EQ:theo2}
	\end{IEEEeqnarray}
where $\rho_1=3L_\text{tr}^2-d_{3}^2$, $\rho_2=\sqrt{d_3^2+24L_\text{tr}^2}$, $z^*_1=H_e[1-\frac{\rho_1}{L_\text{tr}^2}]^{1\over2}$, $z^*_2=H_e\left[1-\frac{1}{8L_\text{tr}^2}\left(d_{3}^2+12L_\text{tr}^2-\rho_2d_{3}\right)\right]$, and $H_e=\frac{1}{2}\sqrt{d_3^2-L_\text{tr}^2}$.
\end{theo}
\begin{IEEEproof}
	The proof is provided in Appendix \ref{App4}.
\end{IEEEproof}
The above theorem suggests that for small  IRSs  operating in the quadratic power scaling regime, the optimal position of the IRS is  close to  Tx or Rx,  which is in agreement with the results for RF-IRS in \cite{RF-OptPos, Marzieh_Poor}. Moreover,  IRSs operating in the linear power scaling regime achieve better performance close to the Tx. 
However, when the IRS size is  large, the optimal position of  the IRS is  equidistant from  Tx and Rx. 
\subsection{Optimal Position of  DF Relay}
For a DF relay-based link, the optimal position of the relay at high SNR is determined by the diversity gain. Thus,  minimizing the outage performance at high SNR in (\ref{EQ:Opt_prob}) is equivalent to maximizing the diversity gain  of the relay-based FSO link, $D_\text{rel}$, and
thus, as shown in  \cite{Sahar_Placement}, the  optimal position of the relay $(x_o^*, z_o^*)$  is  equidistant  from the Tx and Rx and  given by 
\begin{IEEEeqnarray}{rll}
	(x_o^*, z_o^*)=\left(0, H_e\right).
	\label{EQ:Opt_pos_rel}
\end{IEEEeqnarray}
\vspace*{-8mm}
\section{Simulation Results}\label{Sec_Sim}

\begin{figure}[t!]
	\centering
	\includegraphics[width=0.55\textwidth]{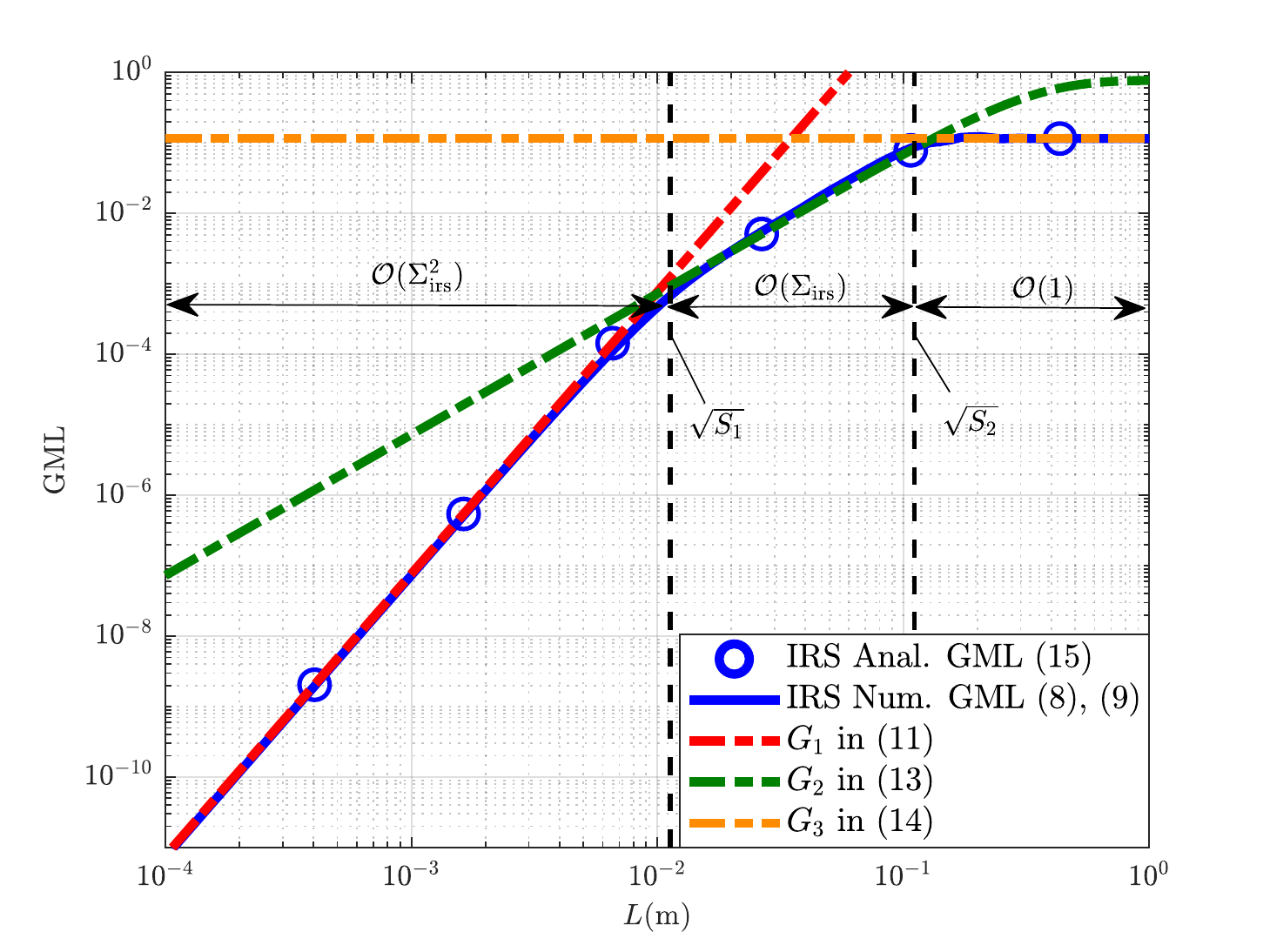}
		\vspace*{-4mm}
		\caption{ GML, $h_{\text{gml},3}$,  vs. IRS length, $L$, for $w_{o1}=2.5$ mm.\label{Fig:GML} }
		\vspace*{-6mm}
\end{figure}

In the following, we consider an FSO system with $\lambda=1550$ nm, $\zeta=1$, noise spectral density $N_0=-114\ \mathrm{dBm}/\mathrm{MHz}$, $\kappa=0.43\times 10^{-3} \frac{\mathrm{dB}}{\mathrm{m}}$, $C_n^2=50\times 10^{-15}$, $\eta=377\,  \Omega$, $P_\text{tot}=1$ mW, $d_{3}=2$ km, $L_\text{tr}=1.6$ km, and $a=10$ cm \cite{IRS_FSO_WCNC}.  We adopt a square-shaped IRS with $L_x=L_y=L$ and assume that the IRS and relay are centered at $(x_o,z_o)=(0,H_e)$, respectively, unless  specified otherwise. 
Fig.~\ref{Fig:GML} shows the GML  of the IRS-assisted FSO link, $h_\text{gml,3}$, versus the length of the IRS, $L$. As can be observed, the numerical GML  in (\ref{gain}) and (\ref{Huygens}) matches the analytical approximation in (\ref{EQ:Theo1}).
 Moreover, depending on the IRS size, the analytical GML in (\ref{EQ:Theo1}) is determined by  ${G}_1$ in (\ref{EQ:ApproxG_1}), ${G}_2$ in (\ref{EQ:ApproxG_2}), and $G_3$  in (\ref{EQ:Lem3}), where the dashed vertical lines indicate the boundary values $\sqrt{S_1}=1.13$ cm, and $\sqrt{S_2}= 0.11$ m.  
  Fig.~\ref{Fig:GML} confirms that for an IRS  length of $L\leq \sqrt{S_1}$,  the GML  ${G}_1$ increases quadratically with the IRS size $L^2$, see (\ref{EQ:ApproxG_1}). By further increasing the IRS length, in the range  $\sqrt{S_1}\leq L \leq \sqrt{S_2}$, the IRS collects the power of the tails of the Gaussian beam incident on the IRS and the GML scales linearly with the IRS size $L^2$. Finally, for IRS lengths of $ L\geq \sqrt{S_2}$, due to the limited  lens size, the IRS-based GML saturates to  $G_3$ in (\ref{EQ:Lem3}). 
 
\begin{figure}[t!]
	\centering
	\includegraphics[width=0.55\textwidth]{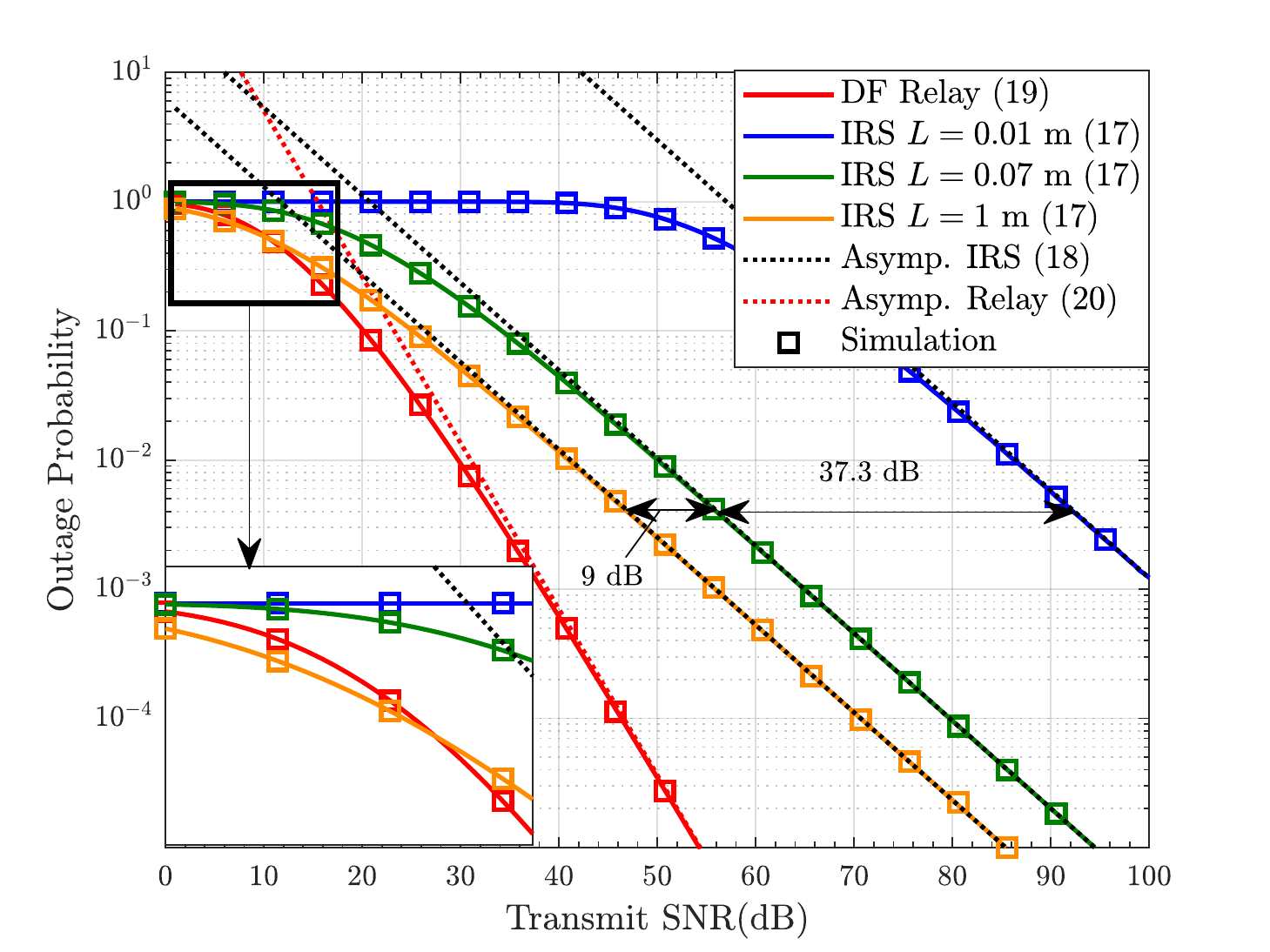}
	\vspace*{-4mm}
	\caption{ Outage probability vs. transmit SNR with $w_{o1}=w_{o2}=7$ mm and  $\gamma_{th}=0$ dB.\label{Fig:Outage}}
	\vspace*{-6mm}
\end{figure}

Fig.~\ref{Fig:Outage} shows the outage probability  of  relay- and IRS-assisted FSO links for IRS lengths of $L=0.01$ m, $0.07$ m, and 1 m and a threshold SNR of $\gamma_{th}=0$ dB versus the transmit SNR, $\bar{\gamma}$.   As can be observed, the analytical outage probabilities for the relay in (\ref{EQ:Outage_relay}) and the IRS in (\ref{EQ:Outage_IRS}) match the simulation results.  Furthermore,  the asymptotic outage probability  for IRS- and FSO-assisted links in (\ref{EQ:Diversity_IRS}) and (\ref{EQ:Diversity_Relay}), respectively, become accurate for high SNR values. As can be observed from Fig.~\ref{Fig:Outage},  due to distance-dependent fading parameters, the diversity gain of the relay-assisted FSO link  is approximately two times larger than that of  the IRS-assisted link, i.e., $\frac{D_\text{rel}}{D_\text{irs}}=\frac{\min\{\varrho_1,\varrho_2\}}{\varrho_3}=1.9$.  Moreover,  by increasing the IRS length from $0.01$ m to $0.07$ m, the FSO link gains 37.3 dB in SNR due to the linear scaling of the received power with the IRS size, see Fig.~\ref{Fig:GML}. However, when the IRS length increases from $0.07$ m to $1$ m, the  received power saturates at a constant value and the additional SNR  gain is only 9 dB. Furthermore, Fig.~\ref{Fig:Outage} reveals that for the adopted system parameters, an IRS with $L=1$ m outperforms the relay at low SNR values, although the performance difference is  small.
 
\begin{figure}
	\centering
	\includegraphics[width=0.55\textwidth]{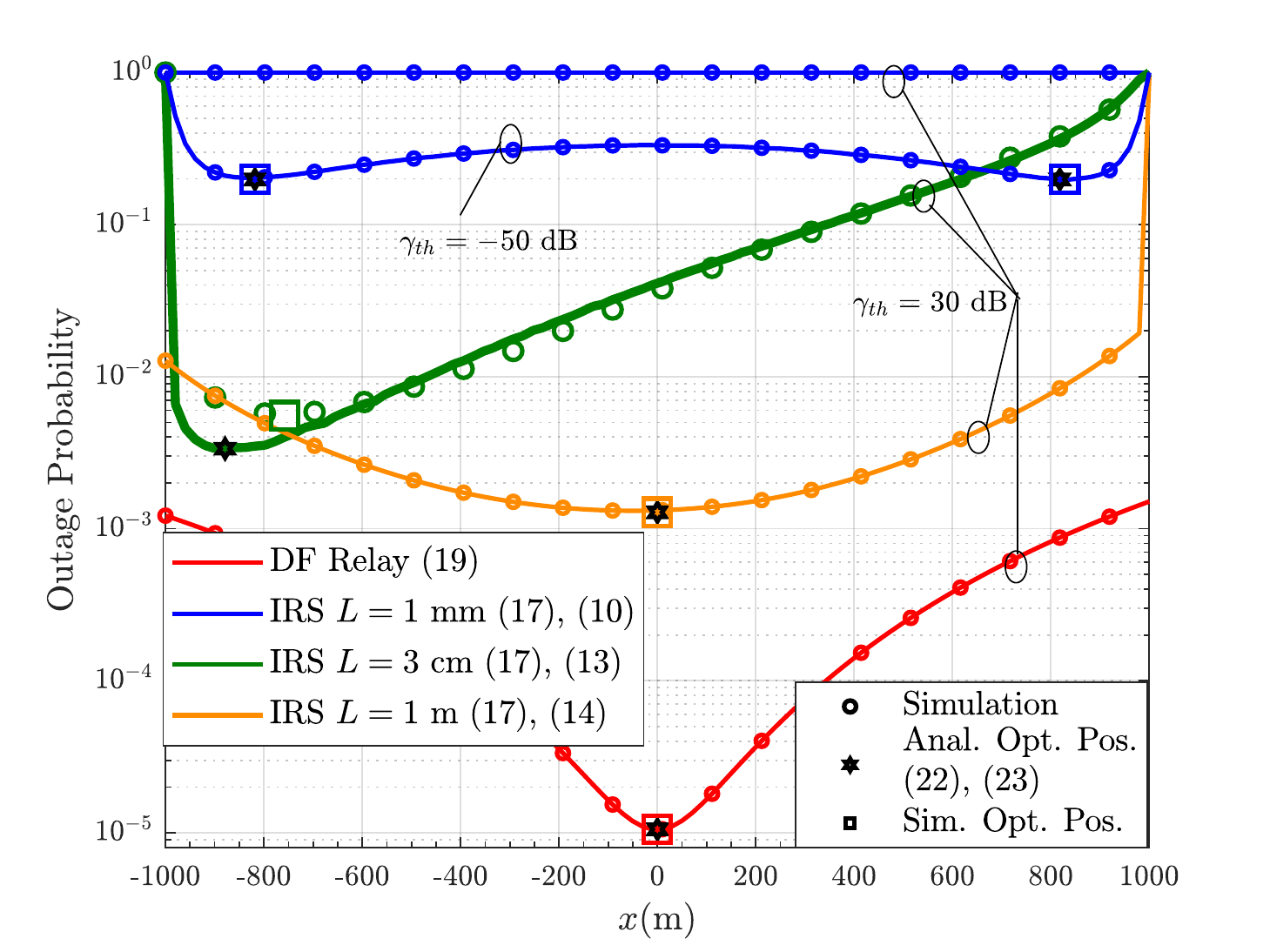}	
	\vspace*{-4mm}
	\caption {Outage probability vs. position of the $x$-coordinate of the center of IRS and relay with $w_{o1}=w_{o2}=7$ mm and $\bar{\gamma}=84$ dB. \label{Fig:Pos}}
	\vspace{-6mm}
\end{figure}

Fig.~\ref{Fig:Pos} shows the outage probability of the IRS- and relay-assisted links for $\gamma_{th}= 30$ dB versus the location of the center of the IRS/relay on the $x$-axis.  To better illustrate the outage performance of extremely small IRSs, we also show results for $\gamma_{th}=-50$ dB. The optimal  positions obtained from the analytical results in (\ref{EQ:theo2}), (\ref{EQ:Opt_pos_rel}) and  simulations are denoted by   \DavidStarSolid\, and $\square$, respectively. The analytical outage probability for   the relay-assisted link in (\ref{EQ:Outage_relay}) matches the simulation results. We obtained the outage probability for the IRS-assisted link in  (\ref{EQ:Outage_IRS})  based on GMLs $G_1$ in (\ref{EQ:Lem1}) , ${G}_2$ in (\ref{EQ:ApproxG_2}), and $G_3$ in (\ref{EQ:Lem3}) for IRS sizes of $L=1$ mm, 3 cm, and 1 m, respectively. The analytical outage performance matches the simulation results except for an IRS length of $L=3$ cm.  The reason is that the IRS with $L=3$ cm does not always operate in the  linear power scaling regime, since the boundary values $S_1$ and $S_2$ in (\ref{EQ:Theo1}) change with the position of the IRS. However,  despite the small discrepancy between  simulation and analytical results for $x<-600$ m, the analytical optimal placement still leads to a close-to-optimal simulated outage performance. Furthermore, as can be observed, the optimal position of the relay is  equidistant from Tx and Rx  which matches the analytical result (\ref{EQ:Opt_pos_rel}). Moreover, for  different IRS sizes, different optimal positions are expected.  For a small IRS length of $1$ mm, the IRS operates in the quadratic power scaling regime and the optimal location is close to the Tx or Rx.  However, when the IRS size is large, i.e., $L=1$ m,  the optimal  position is equidistant from Tx and Rx . For IRSs with length $L=3$ cm, the optimal IRS position  is  close to the Tx as expected from (\ref{EQ:theo2}). 

\vspace*{-3mm}
\section{Conclusions}\label{Sec_concl}
\vspace*{-1mm}
In this paper, we analyzed the power scaling laws for IRS-assisted FSO systems. Depending on the beam waist, position of Tx and Rx w.r.t. the IRS, and lens radius, the received power at the lens grows  quadratically or linearly with the IRS size or remains constant. We analyzed the GML,   the boundary IRS sizes, and the asymptotic outage performance for these power scaling regimes.  Our results show that, at the expense of higher hardware complexity, a relay-assisted link    outperforms an IRS-assisted link at high SNR, although at low SNR the IRS-assisted link can  be beneficial. We also compared the optimal IRS placement for the different power scaling regimes with the optimal relay placement.  IRSs of small size achieve optimal outage   performance close to the Tx (Rx), whereas large size IRSs perform better when  placed equidistant from Tx and Rx.  
\appendices
\renewcommand{\thesectiondis}[2]{\Alph{section}:}
\vspace*{-3mm}
\section{Proof of Lemma~\ref{Lemma1}}\label{App1}
\vspace*{-2mm}
First,  for $L_x\ll w_{\text{in},x}$ and $L_y\ll w_{\text{in},y}$,   the incident Gaussian beam on the IRS can be approximated by a plane wave as $\tilde{E}_\text{in}=\tilde{E}_oe^{-jk(d_1-x_r\cos(\theta_i))}$.
To obtain $\tilde{E}_o$, we ensure that the powers of the plane wave and  the Gaussian beam incident on the IRS are equal. Thus, $	\frac{1}{2\eta}\iint_{\Sigma_\text{irs}} |\tilde{E}_0|^2\mathrm{d}x_r\mathrm{d}y_r= P_\text{in}$,
where $P_\text{in}=\tilde{G}_2 P_\text{tot}$ and $\tilde{G}_2$ is given in (\ref{Lemma2}). This leads to $\tilde{E}_o=\sqrt{\frac{2\eta P_\text{tot} \tilde{G}_2\pi}{L_x L_y}}$.
Then, given that $\Sigma_\text{lens}\ll A_\text{rx}$, we use the Huygens-Fresnel principle in (\ref{Huygens}) with the Taylor series approximation of $d_2\big[1+\frac{|\mathbf{r}_{p2}-\mathbf{r}_r|^2}{d_2^2}\big]^{1\over2}\!\!\approx d_2-\frac{x_r(\sin(\theta_{r})x_{p2}+\cos(\theta_{r}))+y_ry_{p2}}{d_2}$, the reflected electric field is   as follows
\vspace*{-2mm}
\iftoggle{Conf}{
	\begin{IEEEeqnarray}{rll}
		&{E}_r(\mathbf{r}_{p2}) =\tilde{C}\text{sinc}\left(\frac{kL_x \sin(\theta_r)}{2d_2}x_{p2}\right)\text{sinc}\left(\frac{kL_y}{2d_2}y_{p2}\right)\!,\qquad
		\label{EQ:Proof_Lem1_sinc}
\end{IEEEeqnarray}
\vspace*{-2mm}
}
{	
\begin{IEEEeqnarray}{rll}
	{E}_r(\mathbf{r}_{p2})= \tilde{C}  \text{sinc}\left(\frac{kL_x \sin(\theta_r)}{2d_2}x_{p2}\right)\text{sinc}\left(\frac{kL_y}{2d_2}y_{p2}\right),\quad
	\label{EQ:Proof_Lem1_sinc}
\end{IEEEeqnarray}
}
where  $\tilde{C}=L_x L_y\frac{\tilde{E}_o \sqrt{\sin(\theta_r)}}{j\lambda d_2}e^{-jk\left(\Phi_0+d_1+d_2\right)}$.  
Then, we obtain 
\iftoggle{Conf}{
\begin{IEEEeqnarray}{rll}
	&\int \text{sinc}^2(ax) \mathrm{d}x \overset{(a)}{=} \frac{-1}{2a^2x}\left[2ax\text{Si}(2ax)+\cos(2ax)-1\right],\qquad
	\label{EQ:SincInt}
\end{IEEEeqnarray}	
}
{
\begin{IEEEeqnarray}{rll}
	\int \text{sinc}^2(ax) \mathrm{d}x&
	\overset{(a)}{=} \frac{-1}{2a^2x}\left[2ax\text{Si}(2ax)+\cos(2ax)-1\right],
	\label{EQ:SincInt}
\end{IEEEeqnarray}
}
where in $(a)$, we use the partial integration rule. 
Then, we approximate  the circular lens of radius $a$ with a square lens of length $a\sqrt{\pi}$ \cite{Farid_fso} and substitute (\ref{EQ:Proof_Lem1_sinc}) in (\ref{gain}). Then,  by applying (\ref{EQ:SincInt}), we obtain (\ref{EQ:Lem1}) and this completes the proof.
\vspace*{-2mm}
\section{Proof of Lemma~\ref{Lemma2}}\label{App2}
The electric field of the incident Gaussian beam on the IRS, $E_{\text{in}}({\mathbf{r}_r})$, is given by \cite[Eq.(7)]{TCOM_IRSFSO}, thus, the incident power on the IRS, $P_\text{in}=\frac{1}{2\eta}\iint_{\Sigma_\text{irs}}\left|E_{\text{in}}({\mathbf{r}_r})\right|^2 \mathrm{d}\mathbf{r}_r$, is given by
\begin{IEEEeqnarray}{rll}
	&P_\text{in}=\frac{2P_\text{tot}\sin(\theta_{i})}{\pi w^2(d_1)} \int_{-\frac{L_x}{2}}^{\frac{L_x}{2}} e^{-\frac{2x_r^2\sin^2(\theta_i)}{w^2(d_1)}} \mathrm{d}x_r\int_{-\frac{L_y}{2}}^{\frac{L_y}{2}} e^{-\frac{2y_r^2}{w^2(d_1)}} \mathrm{d}y_r.\qquad\label{EQ:proofLem2}
\end{IEEEeqnarray}
Then, we solve the above integral with \cite[Eq.~(2.33-2)]{integral}. As we assume the lens size in this case to be much smaller than the IRS size, the lens receives all the  power incident on the IRS. Thus, the GML in this regime  is obtained by normalizing (\ref{EQ:proofLem2}) with $P_\text{tot}$, which leads to (\ref{EQ:Lem2}). This completes the proof.
\vspace*{-2mm}
\section{Proof of Lemma~\ref{Lemma3}}\label{App3}
The  electric field reflected from the IRS and received at the lens, $E_r(\mathbf{r}_{p2})$,  is given by \cite[Eq.~(16)]{TCOM_IRSFSO}. Substituting $L_x,L_y\to \infty$ in $E_r(\mathbf{r}_{p2})$, the GML  in (\ref{gain}) becomes
\vspace*{-2mm}
\iftoggle{Conf}{
\begin{IEEEeqnarray}{rll}
		&G_3\!\!=\!\!C_3\int_{-\frac{a\sqrt{\pi}}{2}}^{\frac{a\sqrt{\pi}}{2}}\!\!\!\!\! e^{-\frac{k^2\sin^2(\theta_p)x_p^2\mathcal{R}\{b_x\}}{2d_2^2|b_x|^2}} \mathrm{d}x_{p2}\!\!\!\int_{-\frac{a\sqrt{\pi}}{2}}^{\frac{a\sqrt{\pi}}{2}}   e^{-\frac{k^2 y_{p2}^2\mathcal{R}\{b_y\}}{2d_2^2|b_y|^2}}  \mathrm{d}y_{p2},\qquad
\end{IEEEeqnarray}
}
{
\begin{IEEEeqnarray}{rll}
	&G_3\!\!=\!\!C_3\int\limits_{-\frac{a\sqrt{\pi}}{2}}^{\frac{a\sqrt{\pi}}{2}}\!\!\!\!\! e^{-\frac{k^2\sin^2(\theta_p)x_p^2\mathcal{R}\{b_x\}}{2d_2^2|b_x|^2}} \mathrm{d}x_{p2}\!\!\!\int\limits_{-\frac{a\sqrt{\pi}}{2}}^{\frac{a\sqrt{\pi}}{2}}   e^{-\frac{k^2 y_{p2}^2\mathcal{R}\{b_y\}}{2d_2^2|b_y|^2}}  \mathrm{d}y_{p2},\qquad
\end{IEEEeqnarray}
}
where  $C_3=\frac{2P_\text{tot}\sin(\theta_{i})\sin(\theta_r)\pi}{\lambda^2w^2(d_1)d_2^2 |b_x||b_y|}$, $R\{\cdot\}$ denotes the real part of a complex number, $b_x=\frac{\sin^2(\theta_i)}{w^2(d_1)}-\frac{jk}{2}\left(\frac{\sin^2(\theta_i)}{R(d_1)}+{\sin^2(\theta_r)\over d_2}\right)$, and $b_y=\frac{1}{w^2(d_1)}-\frac{jk}{2}\left(\frac{1}{R(d_1)}+{1\over d_2}\right)$. Then, substituting  from \cite[Eq.~(2.33-2)]{integral}, we obtain (\ref{EQ:Lem3}) and this completes the proof.
\vspace*{-2mm}
\section{Proof of Theorem~\ref{Lemma4}}\label{App4}
Depending on the IRS size, the optimal position of the IRS is calculated by approximating $h_{\text{gml},3}$ for each power scaling regime. First, the position of the center of the IRS $(x_o,z_o)$ on the ellipse can be rewritten in terms of $d_1$ and $d_2$ as  follows
\vspace*{-2mm}
\begin{IEEEeqnarray}{rll}
x_o=\frac{d_1^2-(d_2)^2}{2L_\text{tr}}, z_o=H_e\Big[1-\frac{\left(d_1^2-d_2^2\right)^2}{d_{3}^2 L_\text{tr}^2}\Big]^{1/2}.\quad
	\label{proof_Theo2_0}
\end{IEEEeqnarray}
 For $\Sigma_\text{irs}\leq S_1 \vee \Sigma_\text{irs}\leq S_3$, the  GML  is $h_{\text{gml},3}\approx{G}_1$. Then, we substitute in (\ref{EQ:ApproxG_1}), the values of $\sin(\theta_i)=\frac{z_o}{d_1}$,  $\sin(\theta_p)=\frac{z_o}{d_p}$,  $z_o$ given in (\ref{proof_Theo2_0}),  and $d_2=d_3-d_1$.
Next, by solving $\frac{\mathrm{d}{G}_1}{\mathrm{d}d_1}=0$, the extremal points, comprising maxima and minima, are given by
\begin{IEEEeqnarray}{rll}
	&d_{1,\text{min}}^{(1)}= \frac{d_{3}}{2},\,	d_{1, \text{max}}^{(2)}=\frac{d_{3}}{2}+ \frac{\sqrt{2\rho_1}}{4},\,
	&d_{1,\text{max}}^{(3)}=\frac{d_{3}}{2}- \frac{\sqrt{2\rho_1}}{4}.\qquad
	\label{proof_Theo2_2}
\end{IEEEeqnarray}
Then,   substituting the maxima in (\ref{proof_Theo2_0})  leads to (\ref{EQ:theo2}).

Next, for $S_1\leq \Sigma_\text{irs}\leq S_2$, the GML  is $h_{\text{gml},3}\approx{G}_2$. Then, we substitute $\sin(\theta_i)=\frac{z}{d_1}$ and $d_2=d_{3}-d_1$ in (\ref{EQ:ApproxG_2}).
Then, by solving $\frac{{G}_2}{\mathrm{d}d_1}=0$, the extremal points  are given by 
\vspace*{-2mm}
\begin{IEEEeqnarray}{rll}
	d_{1,\text{max}}^{(1)}=\left(5d_{3}+ \sqrt{\rho_2}\right)/8,\quad
	d_{1,\text{max}}^{(2)}=\left(5d_{3}- \sqrt{\rho_2}\right)/8.
\end{IEEEeqnarray}
Here, $d_{1,\text{max}}^{(1)}$ does not lie on  the ellipse, since  $d_{1,\text{max}}^{(1)}>\max(d_1)$, where $\max(d_1)=\frac{d_{3}+L_\text{tr}}{2}$.  Then, substituting $d_{1,\text{max}}^{(2)}$ in (\ref{proof_Theo2_0}) leads to (\ref{EQ:theo2}).

Next, for $\Sigma_\text{irs}>S_2 \vee \Sigma_\text{irs}> S_3 $, the GML  is $h_{\text{gml},3}\approx G_3$. Then, assuming $d_1\gg z_{R1}$, we  substitute $w(d_1)\approx\frac{\lambda d_1}{\pi w_{o1}}$, $R(d_1)=d_1$, and $d_2=d_{3}-d_1$  in (\ref{EQ:Lem3}).
The maximum of the $\text{erf}(\cdot)$ functions in (\ref{EQ:Lem3}) occur for the minimum of the beamwidths $W_{\text{eq}, x}$ and $W_{\text{eq}, x}$. By solving  $\frac{\mathrm{d} W_{\text{eq}, i}}{\mathrm{d}d_1}=0, i\in\{x,y\}$, we obtain similar minimal points at $\frac{d_{3}}{2}$. Thus, both $\text{erf}(\cdot)$ functions in (\ref{EQ:Lem3})  are maximized at $\frac{d_{3}}{2}$, which in turn maximizes $G_3$. Substituting $\frac{d_{3}}{2}$ in (\ref{proof_Theo2_0}) leads to (\ref{EQ:theo2}) and this completes the proof.

\bibliographystyle{IEEEtran}
\bibliography{My_Citation_1-07-2022}
\end{document}